\documentclass[amsmath,amssymb,preprintnumbers,nofootinbib,preprint,a4paper]{revtex4}
\pdfoutput=1
\usepackage{amsthm}
\usepackage{graphicx}
\usepackage{color}
\newcommand{\beq}{\begin{eqnarray}}
\newcommand{\eeq}{\end{eqnarray}}

\newcommand{\bmp}{\noindent\begin{minipage}{16cm}}
\newcommand{\emp}{\end{minipage}\vskip 7mm}

\usepackage{dcolumn}
\usepackage{bm}
\usepackage{bbm}
\usepackage{subfigure}
\usepackage{pxfonts} 
\usepackage{epsfig}
\usepackage[ margin=5pt, font=normalsize,labelfont=bf,justification=raggedright]{caption}
\usepackage{youngtab}
\usepackage{slashed}
\definecolor{rossoCP3}{cmyk}{0,.88,.77,.40}
\newcommand{\be}{\begin{eqnarray}}
\newcommand{\ee}{\end{eqnarray}}

\begin{document}
\title{\Large  \color{rossoCP3} ~~\\  Pseudo Goldstone Bosons Phenomenology \\  in \\ Minimal Walking Technicolor  }
\author{Tuomas {\sc Hapola}$^{\color{rossoCP3}{\varheartsuit}}$}
\email{hapola@cp3.dias.sdu.dk} 
\author{Federico {\sc Mescia}$^{\color{rossoCP3}{\vardiamondsuit}}$}
\email{mescia@ub.edu} 
\author{Marco {\sc Nardecchia}$^{\color{rossoCP3}{\varheartsuit}}$}
\email{nardecchia@cp3.dias.sdu.dk}  
\author{Francesco {\sc Sannino}$^{\color{rossoCP3}{\varheartsuit}}$}
\email{sannino@ cp3.dias.sdu.dk}  
\affiliation{$^{\color{rossoCP3}{\varheartsuit}}${\color{rossoCP3} CP$^{3}$-Origins} \& 
Danish Institute for Advanced Study {\color{rossoCP3} DIAS}, 
University of Southern Denmark, Campusvej 55, DK-5230 Odense M, Denmark,}
\affiliation{$^{\color{rossoCP3}{\vardiamondsuit}}
$Dept. d'Estructura i Constituents de la Mat\`eria and Institut de Ci\`encies del Cosmos, Universitat de
Barcelona, Diagonal 647, E-08028 Barcelona, Spain.\vspace*{1cm}
}

\vspace*{2cm}
\begin{abstract}
We construct the non-linear realized Lagrangian for the Goldstone Bosons associated to the breaking pattern of SU(4) to SO(4).  This pattern is expected to occur in any Technicolor extension of the standard model featuring two Dirac fermions transforming according to real representations of the underlying gauge group.  We concentrate on the  Minimal Walking Technicolor quantum number assignments with respect to the standard model symmetries. We demonstrate that for, any choice of the quantum numbers, consistent with gauge and Witten anomalies the spectrum of the pseudo Goldstone Bosons contains electrically doubly charged states which can be discovered at the Large Hadron Collider.  
\\[.1cm]
{\footnotesize  \it Preprint: CP$^3$-Origins-2012-03 \&  DIAS-2012-04 \& 
ICCUB-12-077 \&ECMUB69}
 \end{abstract}

\maketitle
\newpage

\section{Introduction}
The main physics objective of the Large Hadron Collider (LHC) experiment is to uncover the mechanism responsible for the Electro Weak Symmetry Breaking (EWSB). One fascinating possibility is that the EWSB has a dynamical origin \cite{Weinberg:1979bn,Susskind:1978ms} 
related to a new strongly interacting field theory similar to Quantum Chromo Dynamics. At times this scenario is also referred to as Technicolor (TC). Gauge theories featuring only fermionic degrees of freedom are free from quadratic divergences and therefore offer a natural solution to the standard model (SM) hierarchy problem. Within this scenario, at the EW scale it is not possible to use perturbation theory  because the new underlying gauge coupling is  large. The strong sector can, however, be constrained using its global symmetries. 

We define with $\mathcal{G}$  the new global symmetry which  breaks dynamically to the stability group $\mathcal{H}$. The minimal and most investigated possibility is to consider the pattern of breaking $SU(2)_L \times SU(2)_R \to SU(2)_V$.  $SU(2)_V$ acts as custodial symmetry guaranteeing the relation  $M_W=\cos\theta_w M_Z$ to hold at the tree level. As a consequence of this minimal pattern of breaking all the Goldstone Bosons (GB) become the longitudinal degrees of freedom of the massive $W$ and $Z$ gauge bosons in the unitary gauge.  New massive resonances, classified according to the different representations of the unbroken $SU(2)_V$ symmetry as well as space-time symmetries, will appear. The relative ordering in mass of the new states will depend on the specific underlying strongly interacting theory. In \cite{Andersen:2011yj,Andersen:2011nk,Belyaev:2008yj,He:2007ge,Accomando:2008jh,Barbieri:2008cc,Hirn:2007we,Cata:2009iy} the reader can find the phenomenological implications for colliders of this pattern of chiral symmetry breaking. However, a generic vector-like, strongly coupled gauge theory will typically feature larger global symmetry groups.  In fact, even when considering the most minimal TC theories featuring just two Dirac flavors  three distinct patterns of chiral symmetry breaking emerge \cite{Peskin:1980gc,Preskill:1980mz}. If the fermions belong to a complex representation of the underlying TC gauge group then the expected pattern is the one above, i.e. $SU(2)\times SU(2) \rightarrow SU(2)$, if the underlying representation is pseudoreal then the unbroken symmetry is $SU(4)$ expected to break to $Sp(4)$, finally if the representation is real then one expects $SU(4)$ to break to $SO(4)$. We say that it is expected to break to a given maximal subgroup since it is not a mathematical proof that it breaks to this subgroup \footnote{ However, in a few cases certain patterns have been confirmed. For example for the $SU(2)$ TC theory with two Dirac fermions transforming according to the fundamental representation of the underlying TC gauge group there is a definitive Lattice proof \cite{Lewis:2011zb} that $SU(4)$ breaks to $Sp(4)$. The effective, linear and non-linear, Lagrangians for $SU(4)$ breaking to $Sp(4)$ describing also the interactions with the SM have been constructed in \cite{Duan:2000dy,Appelquist:1999dq,Ryttov:2008xe}.}.

In this work, the following pattern of chiral symmetry breaking
\begin{equation}
SU(4) \rightarrow SO(4) \ ,
\end{equation} 
is considered. This is the first pattern of chiral symmetry breaking for minimal models of TC with a very rich spectrum of Pseudo Goldstone Bosons (PGB)s. Of the 9 PGBs associated, 3 are eaten by the SM gauge bosons while the others are physical and can be observed at LHC.  Our main target is to study the phenomenological implications of these states. Furthermore, at least two distinct extensions of the SM of TC type have been constructed potentially underlying this pattern of chiral symmetry breaking.  One model is the Minimal Walking TC (MWT) in which the two Dirac fermions transform according to the adjoint representation of the underlying $SU(2)$ TC gauge group \cite{Sannino:2004qp,Hong:2004td,Dietrich:2005jn,Evans:2005pu,Foadi:2007ue} and the other is the Orthogonal TC \cite{Frandsen:2009mi,Sannino:2009aw} in which the underlying gauge group is taken to be an orthogonal group and the two Dirac fermions transform according to the vector representation. Both models are expected to be (near) conformal \cite{Dietrich:2006cm,Ryttov:2007cx,Pica:2010mt,Pica:2010xq,Sannino:2009aw}, a propriety which helps alleviating the tension with potentially dangerously large Flavor Changing Neutral Currents. In fact in \cite{Fukano:2010yv} it has been argued that  it is best if the TC theory is already in the conformal window (i.e. displays large distance conformality) and that the extension of TC needed to endow the SM fermions with a mass should, de facto, perturb the conformal theory away from conformality and together with the underlying gauge theory trigger chiral symmetry breaking. This was dubbed ideal walking \cite{Fukano:2010yv}, which is the paradigm we have in mind here.  
	
We will concentrate on the effective theory consisting of the PGBs and new leptons . The latter are required by the internal consistency of the theory.  The remaining heavy particle spectrum  is taken to decouple here.  

After introducing the effective Lagrangian for the PGBs and their interactions with the SM fields as well as the new leptons we press on the phenomenological analysis aimed at their discovery at the LHC. We investigate a number of interesting production and decay channels relevant for the LHC phenomenology while discovering, in the end, a golden channel for the discovery of the MWT PGBs. 

In Section \ref{MWT} we briefly summarize the MWT model, and provide the low energy effective theory for the PGBs in \ref{eMWT} . The coupling to the SM matter is introduced in Section \ref{mMWT}. The PGB production at the LHC is studied in 
Section \ref{PGBproduction} and the decays in \ref{PGBd}. We investigate the PGB discovery potential and conclude in Section \ref{PGBdp}.

The PGB  phenomenological analysis is instrumental in discovering the specific underlying extension of the SM of TC type. A recent analysis of PGBs stemming from traditional TC models in which the techniquarks carry color has been performed in \cite{Chivukula:2011ue}.  We provide a complementary analysis of the phenomenology of PGBs in which the techniquarks do not carry ordinary color and transform according to the real representation of the underlying TC gauge group.

\section{Minimal Walking Technicolor Summary}
\label{MWT}
In the MWT, the extended gauge group is $SU(2)_{TC} \times SU(3)_C \times SU(2)_L \times U(1)_Y$ and the field content of the TC sector is constituted by the techni-fermions, $Q$, $U^c$ and $D^c$, and one techni-gluon all transforming according to the adjoint representation of $SU(2)_{TC}$. 

The model suffers from the Witten topological anomaly \cite{Witten:1982fp} which is cured by
adding a new fermionic weak doublet $L$ singlet under TC \cite{Dietrich:2005jn}.  Furthermore the gauge anomalies cancel when introducing the $SU(2)_L$ singlets $E^c$ and $N^c$ with the hypercharge assignment below\footnote{We use the two component Weyl notation throughout the paper.}:

\begin{minipage}[c]{0.22\textwidth}
\vspace{30pt}
\begin{flushright}
Techniquarks \vspace{40pt}\\ 
New Leptons
\end{flushright}
\end{minipage}\begin{minipage}[c]{0.75\textwidth}
\begin{equation}
\begin{array}{|c|c|c|c|c|}
\hline
\textrm{Field} & SU(2)_{TC} & SU(3)_C & SU(2)_L & U(1)_Y \\
\hline
Q= \left( \begin{array}{c} U \\ D \end{array}\right)    & \mathbf{3} & \mathbf{1} & \mathbf{2} & \frac{y}{2}\\
U^c & \mathbf{3} & \mathbf{1} & \mathbf{1} & -\frac{y+1}{2}\\
D^c & \mathbf{3} & \mathbf{1} & \mathbf{1} & -\frac{y-1}{2}\\
\hline
L     & \mathbf{1} & \mathbf{1} & \mathbf{2} & - \frac{3y}{2}\\
N^c & \mathbf{1} & \mathbf{1} & \mathbf{1} & \frac{3y-1}{2}\\
E^c & \mathbf{1} & \mathbf{1} & \mathbf{1} & \frac{3y+1}{2}\\
\hline
\end{array}
\end{equation}
\end{minipage}\\
\vspace{10pt}

\noindent The parameter $y$ can take any real value \cite{Dietrich:2005jn}. We refer to the states $L$, $E^c$ and $N^c$ as the New Leptons. The condensate which correctly breaks the electroweak symmetry\footnote{A detailed discussion regarding the vacuum alignment for MWT can  be found in \cite{Dietrich:2009ix}. } is $\langle UU^c + DD^c \rangle $. To discuss the symmetry properties of the theory it is
convenient to  arrange the technifermions as a column vector, transforming according to the fundamental representation of $SU(4)$
\beq \hat{Q} = \begin{pmatrix}
U \\
D \\
U^c \\
D^c
\end{pmatrix},
\label{SU(4)multiplet} \eeq
The breaking of $SU(4)$ to $SO(4)$ is driven by the
following condensate \beq \langle \hat{Q}^T E \hat{Q} \rangle \eeq
The matrix $E$ is a $4\times 4$ matrix defined in terms
of the 2-dimensional unit matrix as
 \beq E=\left(
\begin{array}{cc}
0 & \mathbbm{1} \\
\mathbbm{1} & 0
\end{array}
\right) \ . \eeq
The above condensate is invariant under an $SO(4)$ symmetry.

\section{Minimal Walking and its Nonlinear Realization}
\label{eMWT}
The symmetry breaking pattern of the MWT model is $SU(4) \to SO(4)$. 
This leaves us with nine broken  generators with associated GBs.
The resulting low energy effective theory can be organized in a derivative expansion with cut-off scale $ 4\pi F$, where $F$ is the GBs decay constant.  We first introduce the matrix: 
\beq
\label{exp}
U=\exp\left({i\frac{\sqrt{2}}{F}}  \Pi^a X^a \right) \, E \ ,
\eeq
where $\Pi^a$ are the 9 Goldstone bosons and $X^a$ are the 9 broken generators (see Appendix \ref{generators}). $U$ transforms under $SU(4)$ in the following way:
\begin{equation}
U \rightarrow g U g^{T}\  ,  \qquad  g \in SU(4). 
\end{equation}
The leading term appearing in the Lagrangian is:
\beq 
\label{nonlinearkinetic}
L_U = \frac{F^2}{2} {\rm Tr}\left[D_\mu UD^\mu U^\dag\right] \, .  \eeq
With $D$ the electroweak covariant derivative:
\begin{eqnarray}
D_{\mu}U =\partial_{\mu}U - i \left[G_{\mu} U + UG_{\mu}^T\right]  \
, 
\end{eqnarray}
where
\begin{eqnarray}
G_{\mu} & = & g_2 \ W^a_\mu \ L^a + g_1\ B_\mu \left(-{R^3}^T+\sqrt{2}\ y \, S^4\right) \ .
\end{eqnarray}
Here $g_1$ and $g_2$ are the hypercharge and the weak couplings, respectively. The value of $F$ is fixed in order to reproduce correctly the electroweak symmetry breaking $F=\frac{2 m_W }{g_2}$.

\begin{figure}[htbp]
\begin{center}
\includegraphics[width=0.5\textwidth]{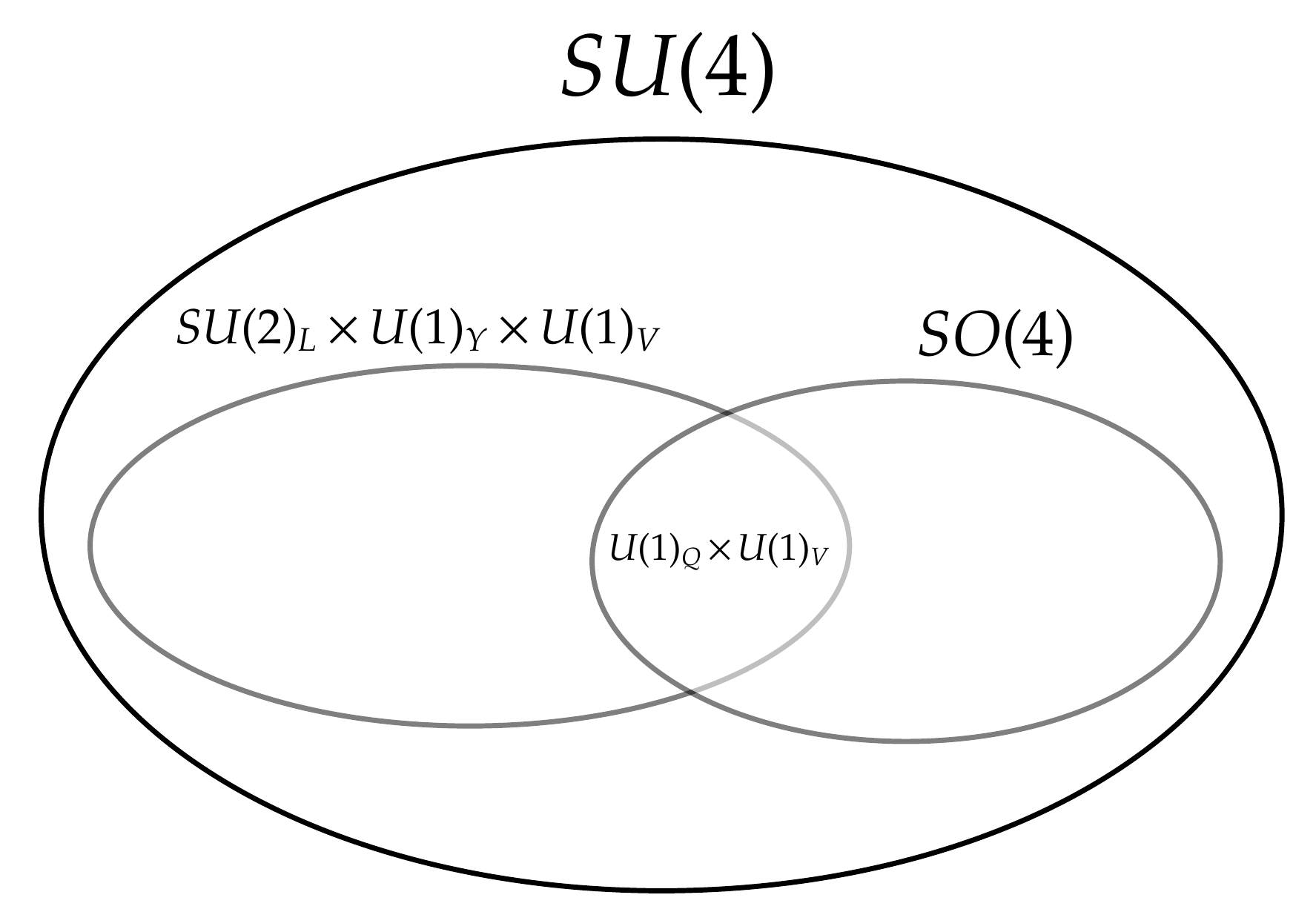}
\caption{Spontaneous and explicit breaking of the $SU(4)$ symmetry.}
\label{breaking}
\end{center}
\end{figure}

The gauging of the electroweak interactions breaks explicitly the $SU(4)$ symmetry group down to $SU(2)_L \times U(1)_Y \times U(1)_V$, while the spontaneous symmetry breaking leaves invariant an $SO(4)$ subgroup.
The remaining unbroken group is $U(1)_Q \times U(1)_V$. The $U(1)_Q$ factor is the symmetry group associated to the electromagnetism while the $U(1)_V$ leads to the conservation of the technibarion number. A simple illustration of the spontaneous and explicit breaking of the $SU(4)$ symmetry is presented in Fig. \ref{breaking}. 

Among the 9 physical degrees of freedom, 3 are eaten up by the longitudinal components of the SM gauge bosons while  the remaining 6 GBs carry technibaryon number and will be denoted by $\Pi_{UU},  \Pi_{UD}$ and $\Pi_{DD}$. Because GB carry technibaryon number, we refer to these states also as technibaryons. 

The bosons can be classified according to the unbroken group $U(1)_{V} \times U(1)_Q $ in the following way:

\begin{equation}
\label{tablepions}
\begin{array}{|c|c|c|c|}
\hline
{\rm Boson} & {\rm U(1)}_{V}  \textrm{ charge} & {\rm U(1)}_{Q}  \textrm{ charge} & \textrm{Linear Combination} \\
\hline 
W^+_L & 0 & +1 & \frac{\Pi^{1} - i \Pi^2}{\sqrt{2}} \\
W^-_L & 0& -1 & \frac{\Pi^{1} + i \Pi^2}{\sqrt{2}} \\
Z_L & 0 & 0 & \Pi^3 \\
\hline
\Pi_{UU} & +1 & y-1 & \frac{\Pi^4+i\Pi^4+\Pi^6+i\Pi^7}{2} \\
\Pi_{DD} & +1 & y+1& \frac{\Pi^4+i\Pi^4+\Pi^6+i\Pi^7}{2} \\
\Pi_{UD} & +1 & y & \frac{\Pi^8+ i \Pi^9}{\sqrt{2}} \\
\hline
\Pi_{UU}^{\dagger} & -1 & -y+1 & \frac{\Pi^4-i\Pi^4+\Pi^6-i\Pi^7}{2} \\
\Pi_{DD}^{\dagger} & -1 & -y-1& \frac{\Pi^4-i\Pi^4+\Pi^6-i\Pi^7}{2} \\
\Pi_{UD}^{\dagger} & -1 & -y & \frac{\Pi^8-i \Pi^9}{\sqrt{2}} \\
\hline
\end{array}
\end{equation}

Electroweak interactions split the technipions masses according to the following pattern~\cite{Dietrich:2009ix}: 
\begin{eqnarray}
\label{splitting}
\Delta m ^2_{\Pi_{UU}} = \frac{m^2_{\textrm{walk}}}{g_1^2+g_2^2} \left[ g_1^2 (1+2y)^2 +g_2^2 \right] \\
\Delta m ^2_{\Pi_{DD}} = \frac{m^2_{\textrm{walk}}}{g_1^2+g_2^2} \left[ g_1^2 (4y^2-1) +g_2^2 \right]  \\
\Delta m ^2_{\Pi_{UD}} = \frac{m^2_{\textrm{walk}}}{g_1^2+g_2^2} \left[ g_1^2 (1-2y)^2 +g_2^2 \right] .
\end{eqnarray}
In models with walking dynamics the value of $m_{\textrm{walk}}$  can be even of few hundreds GeV~\cite{Dietrich:2009ix}. Furthermore it is also possible to introduce  a common  mass term  for the  Pseudo GBs   (PGBs) by adding the following term to the Lagrangian \eqref{nonlinearkinetic}:
\begin{eqnarray}
\label{etcmass}
- \frac{m^2_{\textrm{etc}} F^2 }{4}{\rm Tr} \left[U^{\dagger}B_VUB_V \right]
\end{eqnarray}
with
\begin{eqnarray}
B_V= \left(
\begin{array}{cc}
\mathbbm{1} & 0 \\
0 & -\mathbbm{1}
\end{array}
\right) \ .
\end{eqnarray}
This term was already added in \cite{Foadi:2007ue} and it is expected to emerge from a more complete theory of SM fermion mass generation. It is expected to emerge from a four-techniquark interaction term, and preserves $SU(2)_L \times SU(2)_R \times U(1)_V$ of $SU(4)$, which contains the $SU(2)_V$ custodial symmetry group.

\section{Coupling of the Minimal Walking Goldstone bosons to Matter}
\label{mMWT}

A complete Extended TC \cite{Eichten:1979ah,Dimopoulos:1979es} (ETC) model would fix all the interactions between the PGBs and the fermions (both the SM and the new leptons). Here, we are not going to study a specific ETC mode but consider the couplings respecting the SM symmetries.

It will be useful to know the transformation properties of the matrix $U$ with respect to the electroweak gauge group. Under $SU(4)$, the matrix $U$ transforms as a two index symmetric tensor, or in other words $U$ transforms  like the irreducible representation $\mathbf{10}$ of $SU(4)$:
\begin{equation}
U \rightarrow g U g^{T}\  ,  \qquad  g \in SU(4). 
\end{equation}

Knowing the embedding of the electroweak generators in the $SU(4)$ algebra (see the Appendix \ref{generators} for the details) it is possible to decompose  the representation $\mathbf{10}$ according to the $SU(2)_L \times U(1)_Y$ group:
\begin{eqnarray}
\label{ewqnumbers}
\mathbf{10} &\to& \mathbf{3}_{y} + \mathbf{2}_{1/2}  + \mathbf{2}_{-1/2} + \mathbf{1}_{-y+1} + \mathbf{1}_{-y}   + \mathbf{1}_{-y-1} \ .  
\end{eqnarray}
 The identification of these representations inside the $U$ matrix is given by:
\begin{equation}
\begin{array}{cc}
\begin{array}{ccc}
\mathbf{3}_{y} &\to& T_L \\
\mathbf{2}_{1/2} &\to & H_2 \\
\mathbf{2}_{-1/2} &\to& H_1 \\
\mathbf{1}_{-y-1} &\to& S_1 \\
\mathbf{1}_{-y} &\to& S_2  \\
\mathbf{1}_{-y+1} &\to& S_3
\end{array}
&
\qquad
U=
\displaystyle\frac{1}{F}
\left(
\begin{array}{cc}
T_L & 
\begin{array}{cc}
\frac{H_1}{\sqrt{2}} &
\frac{H_2}{\sqrt{2}} 
\end{array}
 \\
\begin{array}{c}
\frac{H_1^T}{\sqrt{2}} \\
\frac{H_2^T}{\sqrt{2}} 
\end{array} & 
\begin{array}{cc}
S_1 & \frac{S_2}{\sqrt{2}} \\
\frac{S_2}{\sqrt{2}} & S_3 
\end{array}
\end{array}
\right)
\end{array}
\end{equation}

By expanding  the exponential in (\ref{exp}) to the second order in the number PGBs fields, we identify the following phenomenologically relevant quantities entering the interaction Lagrangian relevant for the first LHC searches:
\begin{equation}
T_L=
i
\begin{pmatrix}
 \Pi_{UU} & \frac{\Pi_{UD}}{\sqrt{2}} \\
\frac{\Pi_{UD}}{\sqrt{2}} & \Pi_{DD}
\end{pmatrix} \qquad S_1=i \Pi_{UU}^{\dagger} \qquad S_2=i \Pi_{UD}^{\dagger} \qquad S_3=i \Pi_{DD}^{\dagger}
\end{equation}

\begin{equation}
\frac{H_1}{\sqrt{2} \, F} =
 \begin{pmatrix}
1 - \frac{\Pi_{UD}^{\dagger} \Pi_{UD}+2 \, \Pi_{UU}^{\dagger} \Pi_{UU} }{4 \, F^2} \\
- \frac{\Pi_{UD}^{\dagger} \Pi_{DD}+ \Pi_{UU}^{\dagger} \Pi_{UD} }{2 \sqrt{2} \, F^2} 
\end{pmatrix}
\qquad
\frac{H_2}{\sqrt{2} \, F} =
 \begin{pmatrix}
 - \frac{\Pi_{DD}^{\dagger} \Pi_{UD}+ \Pi_{UD}^{\dagger} \Pi_{UU} }{2 \sqrt{2} \, F^2} \\
1 - \frac{\Pi_{UD}^{\dagger} \Pi_{UD}+2 \, \Pi_{DD}^{\dagger} \Pi_{DD} }{4 \, F^2} 
\end{pmatrix} \ . \label{H1H2}
\end{equation}
 
Depending on the specific choice of the hypercharge assignment $y$, one can construct different Yukawa-type interactions involving the matrix $U$.  Odd integer values are required for $y$ to avoid stable composite states with fractional electric charge\footnote{Bound states made by one technifermion and one technigluon have electric charge $-\frac{y \pm 1}{2}$}.  In general electrically charge stable states are excluded by cosmology, see for example discussion in~\cite{Fairbairn:2006gg}.

Within this setup the allowed Yukawa terms are\footnote{Conventions regarding SM quantum numbers are reported in Appendix \ref{SMquantum}}: 
 \begin{eqnarray}
 \label{yukawastart}
{\cal L}_{2HDM} &=& H_2 q u^c + H_1 q d^c + H_1  {l} e^c + H_2 L N^c + H_1 L E^c   \\
\nonumber
&+ & H^c_1 q u^c +  H^c_2 q d^c +  H^c_2  {l} e^c +  H^c_1 L N^c +  H^c_2 L E^c + \textrm{ h.c.} \\
{\cal L}_{\Pi \psi \psi} & = & 
\begin{cases}
y=-3 &  \qquad  S_3^{\dagger} e^c e^c + \textrm{ h.c.}\\
y=-1& \qquad  S_1^{\dagger} e^c e^c + S_2  {l} {l} + T_L^{\dagger} 
 {l} {l} + \textrm{ h.c.}\\
y=+1 & \qquad  S_3 e^c e^c  +  S_2^{\dagger}  {l} {l} + T_L  {l} {l} + \textrm{ h.c.} \\
y=+3  & \qquad S_1 e^c e^c + \textrm{ h.c.}\\
\end{cases}\label{eqpieec}
\\
{\cal L}_{\Pi L L} & = &
\begin{cases}
y=-1 &  \qquad S_1 E^c E^c  + \textrm{ h.c.}\\
y=+1 &  \qquad S_3 N^c N^c + \textrm{ h.c.}
\end{cases}
\\
 \label{yukawaend}
{\cal L}_{\Pi L \psi} & = &
\begin{cases}
y=-5 &  \qquad S_1 E^c e^c + \textrm{ h.c.} \\
y=-3 &  \qquad S_1^{\dagger} L  {l} + S_1 N^c e^c + S_2 E^c e^c + \textrm{ h.c.}  \\
y=-1 &  \qquad S_2^{\dagger} L   {l} + S_2 N^c e^c + T_L^{\dagger} L  {l} + \textrm{ h.c.}  \\
y=+1 & \qquad  H^c_1 L e^c + H_2 L e^c + H_1 N^c   {l} + H_2^c N^c   {l} + S_3 L   {l} + S_3 N^c e^c  + \textrm{ h.c.}  \\
\end{cases}
\end{eqnarray}
$H^c_i$ is defined as $H^c_i \equiv - i \sigma_2 H_i^* $,  {and the 
$e^c$ and $l$ symbols stand respectively for $SU(2)_L$ singlet and doublets states of the $3$ lepton families 
$(e,\mu,\tau)$}. 
In the equations \eqref{yukawastart}-\eqref{yukawaend},  for simplicity, 
we omitted the appropriate $SU(2)_L$ contractions, the flavor indices and the coupling 
constants in front of each term. 
 The interactions in~\eqref{yukawastart}-\eqref{yukawaend} 
are valid for generic ETC models.  Specific models can provide further symmetries which can be exploited to reduce the number of operators in~\eqref{yukawastart}-\eqref{yukawaend}. 
Moreover, it is worth noticing that quarks can couple, to this order, 
only to the Higgs sector { (via ${\cal L}_{2HDM}$)}, 
i.e. quadratic in the number of technipions. 
{ On the other hand leptons,  due to  ${\cal L}_{\Pi \psi\psi}$, couple with a single technipion. 
In Section~\ref{PGBd} we will further exploit and investigate the  r\^ole of ${\cal L}_{\Pi \psi\psi}$ 
 to single out the golden signatures 
allowing to detect the MWT technipions at LHC.}

The allowed values for $y$ can be further restricted by requiring that the lightest state, 
among the PGBs and the new leptons, is electrically charged and stable. According to this, we discard the cases $y=\{-5,3\}$ and the surviving values of  $y$ are $\{-3,-1,1\}$.

\section{PGB production at the LHC}
\label{PGBproduction}
 The LHC has pushed particle physics to a new era after starting to explore physics at the TeV scale. In the absence of discovery, the LHC is imposing strong 
 constraints on several BSM models \cite{ATLAS,CMS}. In oder to understand the potential of the LHC to discover or to set limits on the PGBs sector of MWT, it is crucial to know their production and decay mechanisms. This is the goal of this section. 

We start by recalling that  the $U(1)_V$ is a symmetry of the techincolor theory in isolation and transforms the low energy effective fields as follows: 
 \begin{eqnarray}
\label{uone}
\Pi_{X} & \to & e^{i \alpha} \Pi_{X} , \quad X = \{ UU, UD, DD \} \nonumber \\
\Pi_{X}^{\dagger} & \to & e^{-i \alpha} \Pi_{X}^{\dagger} , \quad X = \{ UU, UD, DD \} \\
\psi_{Y} & \to  & \psi_Y,  \quad Y = \{ \textrm{SM fermions, New Leptons} \} \nonumber
\end{eqnarray}
 Once the TC sector is coupled to the SM matter, and new leptons, such a symmetry  in general breaks. However it is straightforward to show that the $U(1)_V$ symmetry is accidentally conserved in $\mathcal{L}_U + \mathcal{L}_{2HDM}$.
 This leads to the phenomenological relevant consequence that the PGBs can only be produced in pairs. This is so since they have to be produced either via SM gauge bosons or quarks. These couples only via $\mathcal{L}_U + \mathcal{L}_{2HDM}$.
This is not the case with other TC models, where a single technipion production is possible, see for example  \cite{Chivukula:2011ue, Farhi:1980xs,Casalbuoni:1998fs,Lane:1991qh,Lane:1999uh,Manohar:1990eg}.

Furthermore the MWT technifermions are colorless suppressing their production compared to TC models with colored techniquarks. In fact, when the technifermions carry ordinary color, also the production of color singlet states is enhanced via loops of technifermions in gluon initiated processes \cite{Chivukula:2011ue}.
These observations should help understanding the following MWT PGBs production mechanisms:

\begin{itemize}
\renewcommand{\labelitemi}{$\diamondsuit$}
\item Drell-Yan production
\end{itemize}

\begin{figure}[htb!]
\centering
\includegraphics[width=0.2\textwidth]{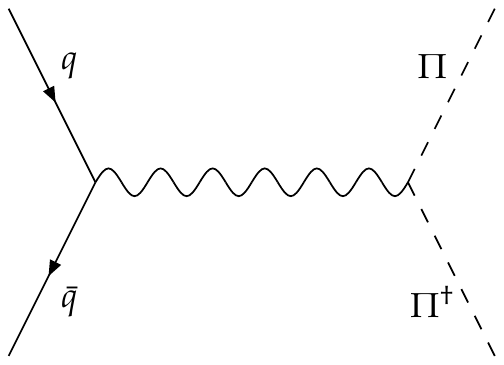}\hspace{0.5cm}\includegraphics[width=0.2\textwidth]{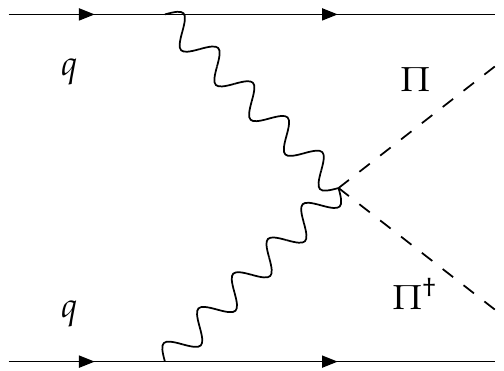}\hspace{0.5cm}\includegraphics[width=0.2\textwidth]{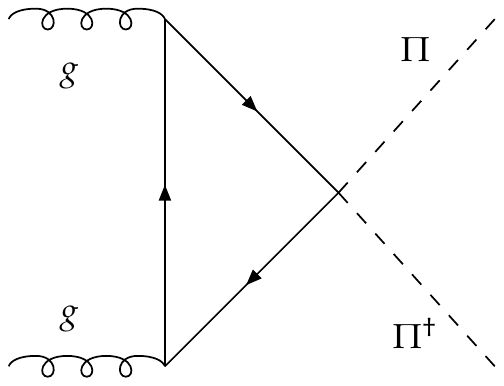}\\
(a)\hspace{100pt} (b)\hspace{100pt}  (c)
\caption{Feynman diagrams for different production mechanisms.}
\label{fig:drell}
\end{figure}

One possibility to produce a PGB pair is through the SM gauge bosons in the s-channel. We take the masses of the PGBs to be above the weak gauge boson masses  and therefore the intermediate gauge bosons cannot be on-shell. The elementary process is shown in Fig.~{\ref{fig:drell}}~(a).

\begin{itemize}
\renewcommand{\labelitemi}{$\diamondsuit$}
\item Production in association with two jets
\end{itemize}

This production can happen through a large number of different diagrams; among which Vector Boson Fusion (VBF) is the most important one. The associated Feynman diagram for VBF is shown in Fig.~{\ref{fig:drell}} (b). One of the advantages presented by this process is the possibility of tagging the two jets coming from the associated quarks, thus reducing the background. 

\begin{itemize}
\renewcommand{\labelitemi}{$\diamondsuit$}
\item Gluon fusion 
\end{itemize}

An initial state with two gluons can produce a PGB pair through a top quark loop. The associated Feynman diagram is shown in Fig.~{\ref{fig:drell}}~(c). The effective Lagrangian describing the gluon fusion process is given in Appendix \ref{eff-coup}. Here we simply derive the coupling between the top-quarks and PGBs. This coupling can be extracted from ${\cal L}_{2HDM}$~\eqref{yukawastart} by expanding $H_1$ and $H_2$ to the second order in the PGB fields (see \eqref{H1H2}), namely
\begin{equation}
\begin{split}
\mathcal{L}_{2HDM,t}
=&\left((y_{1}+y_{2})F-\frac{(y_{1}+y_{2})}{2F}\Pi_{UD}^{\dagger}\Pi_{UD}
-\frac{y_{1}}{2F}\Pi_{UU}^{\dagger}\Pi_{UU}-\frac{y_{2}}{2F}\Pi_{DD}^{\dagger}\Pi_{DD}\right)tt^{c}+ \textrm{ h.c.}
\end{split}
\label{eq:top}
\end{equation}
The mass of the top quark is now $m_{t}=F (y_{1}+y_{2})$. Thus, in general, only one of the PGB pairs couples to the top with a coupling proportional to its mass. The relative size of the couplings $y_{1}$ and $y_{2}$ is unknown and, for simplicity, we choose them to be equal. 

\bigskip
In order to estimate the production cross sections, we implemented the model into Madgraph 5 \cite{arXiv:1106.0522} using FeynRules \cite{arXiv:0806.4194}.  The analyses performed below are done using the CTEQ6L parton distribution functions \cite{Pumplin:2002vw} and we have not implemented any kinematical cuts for the final state PGBs. 

\begin{figure}[htb]
\centering
\includegraphics[width=0.5\textwidth]{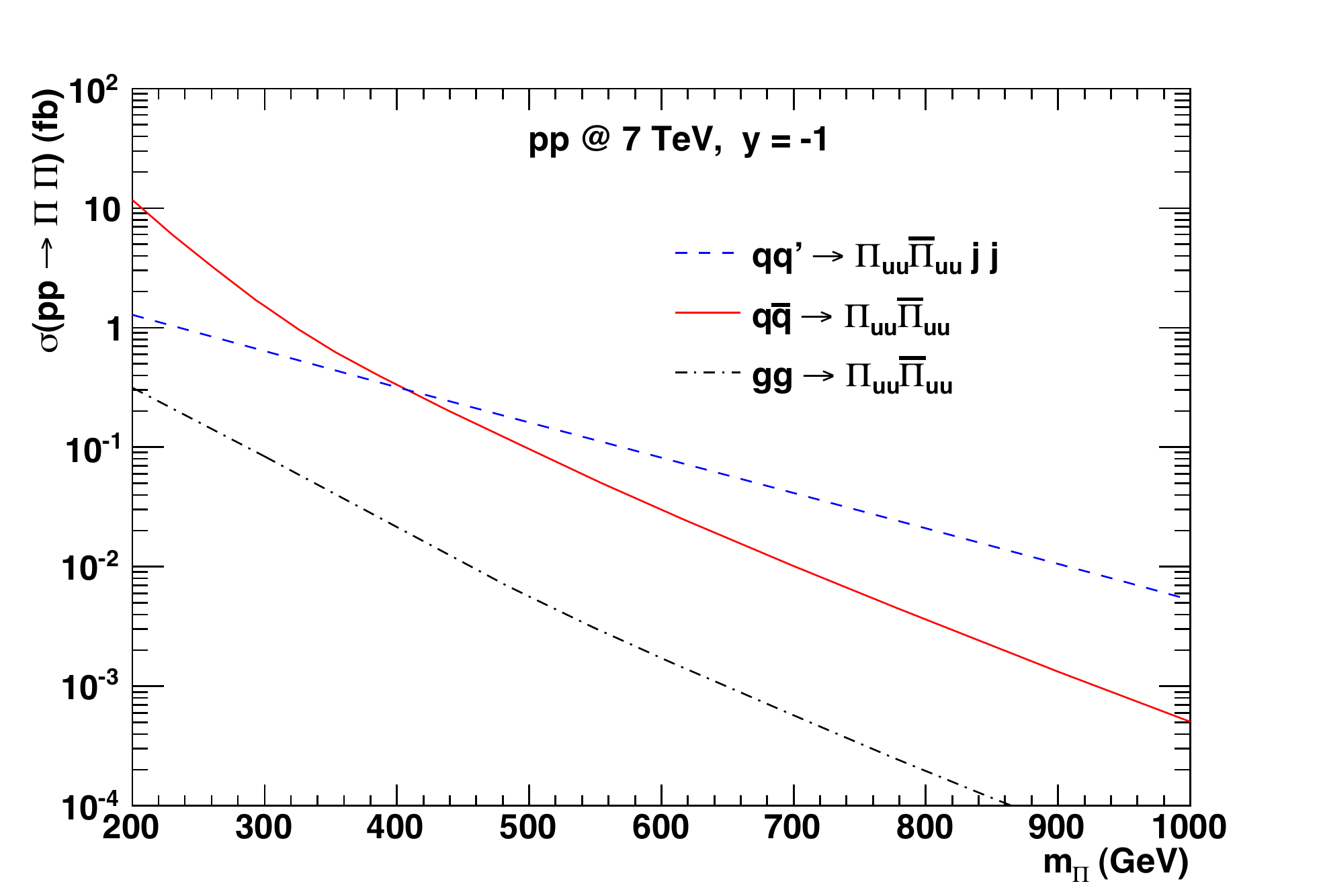}\includegraphics[width=0.5\textwidth]{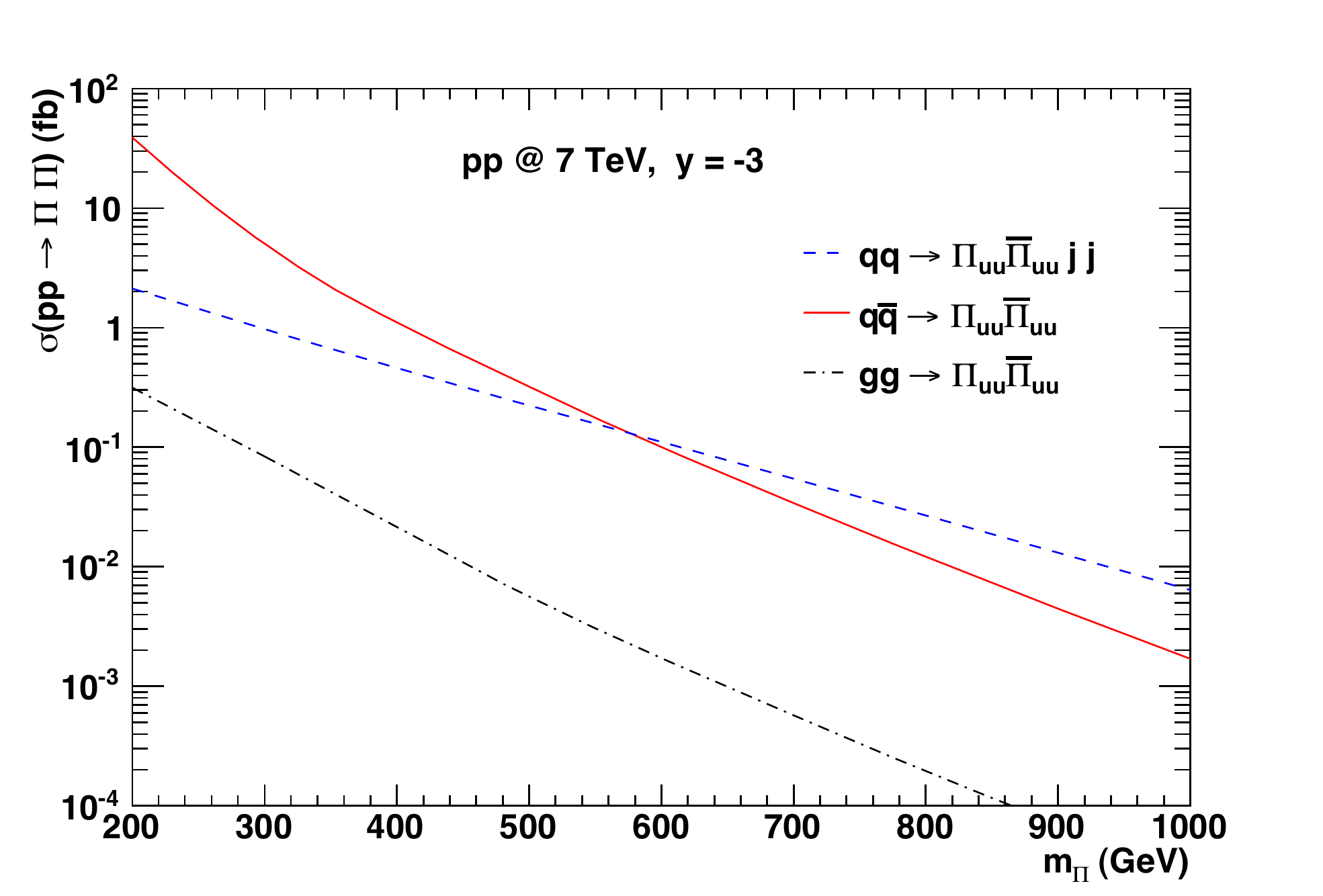}
\includegraphics[width=0.5\textwidth]{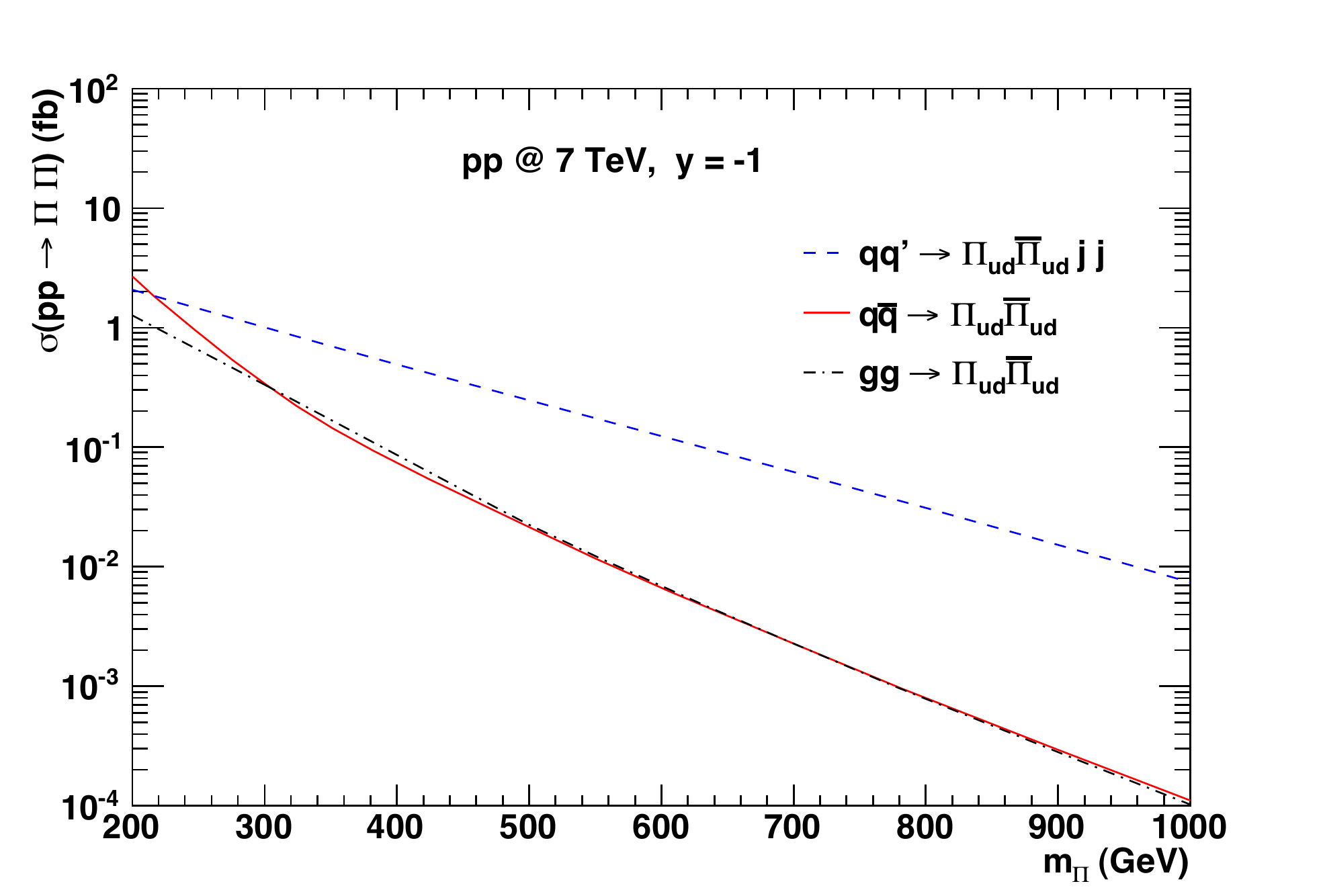}\includegraphics[width=0.5\textwidth]{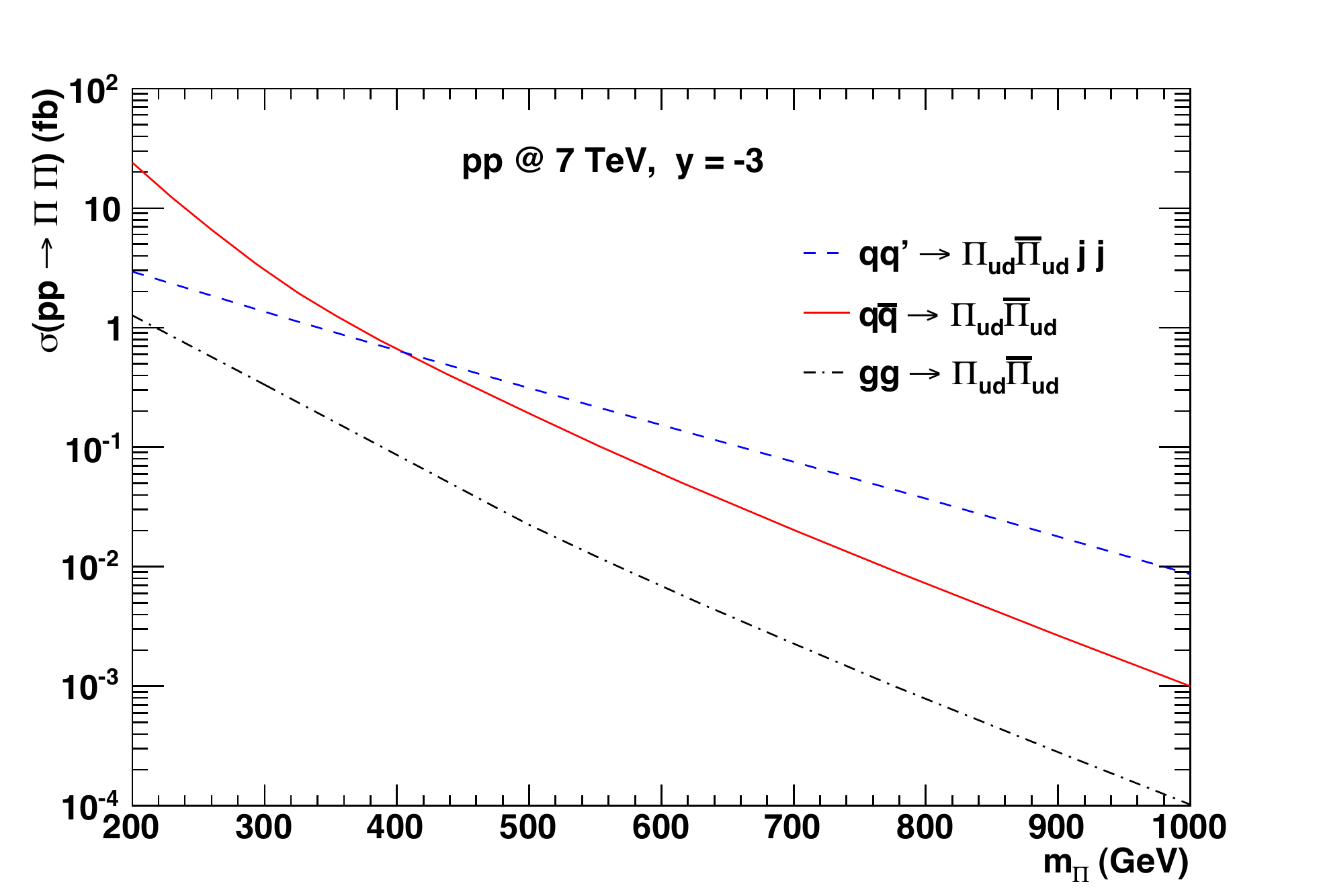}
\includegraphics[width=0.5\textwidth]{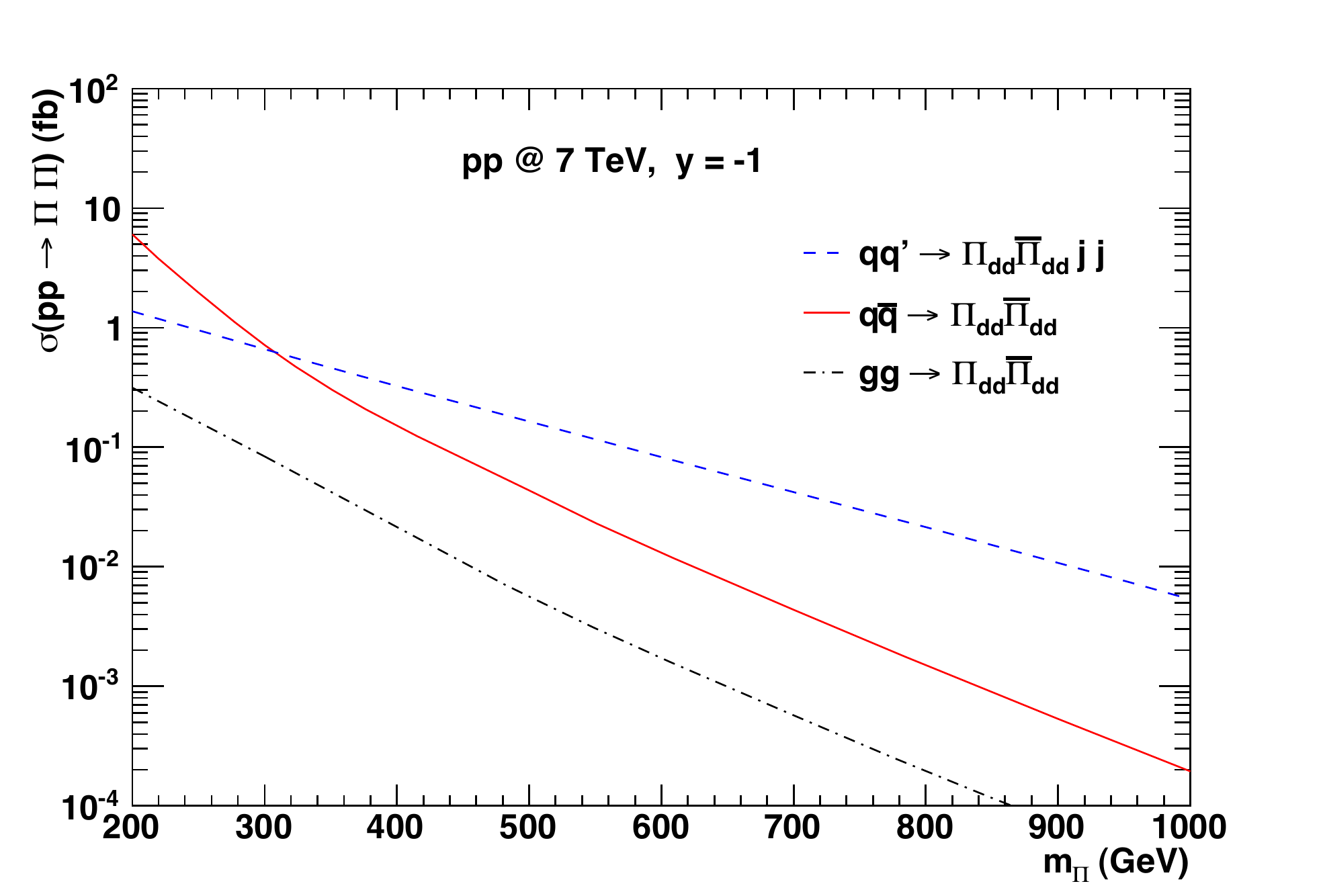}\includegraphics[width=0.5\textwidth]{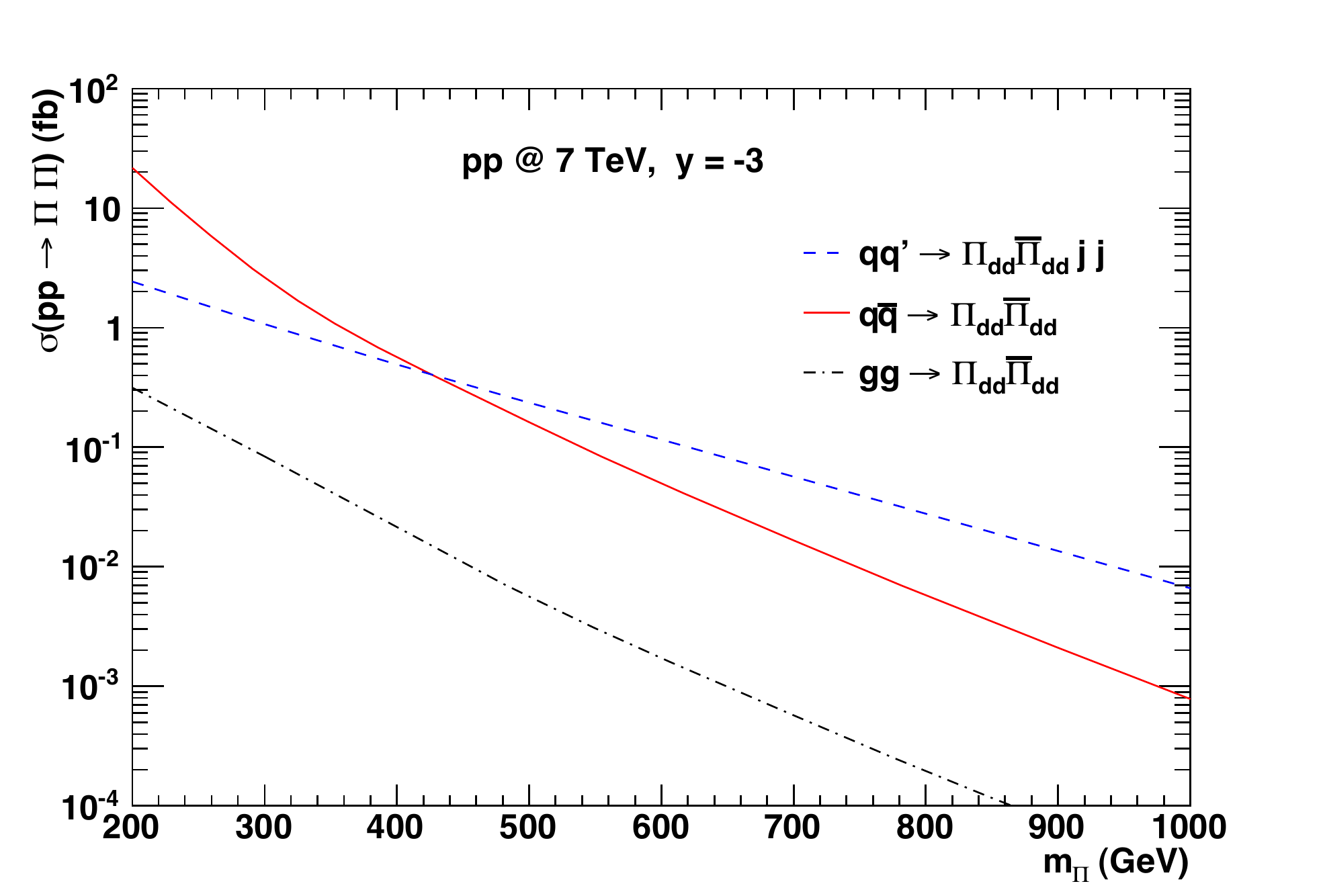}
\caption{The total cross sections for PGB production at the LHC through neutral gauge bosons as a function of the $m_{\Pi}$. The left column is for $y=-1$ and the right column for $y=-3$.} 
\label{fig:y1}
\end{figure}

In Fig. \ref{fig:y1}, the inclusive production cross sections for the PGB pairs with opposite charges are shown  as a function of the PGB mass.  The production cross sections for $y=1$ can be obtained from the case $y=-1$ by interchanging the names of $\Pi_{UU}$ and $\Pi_{DD}$.  For heavier PGBs masses the associate production with a jet pair is the dominant mode whereas in the low mass region 
the proximity of the gauge boson propagator pole enhances the Drell-Yan production.

\begin{figure}[htb]
\centering
\includegraphics[width=0.5\textwidth]{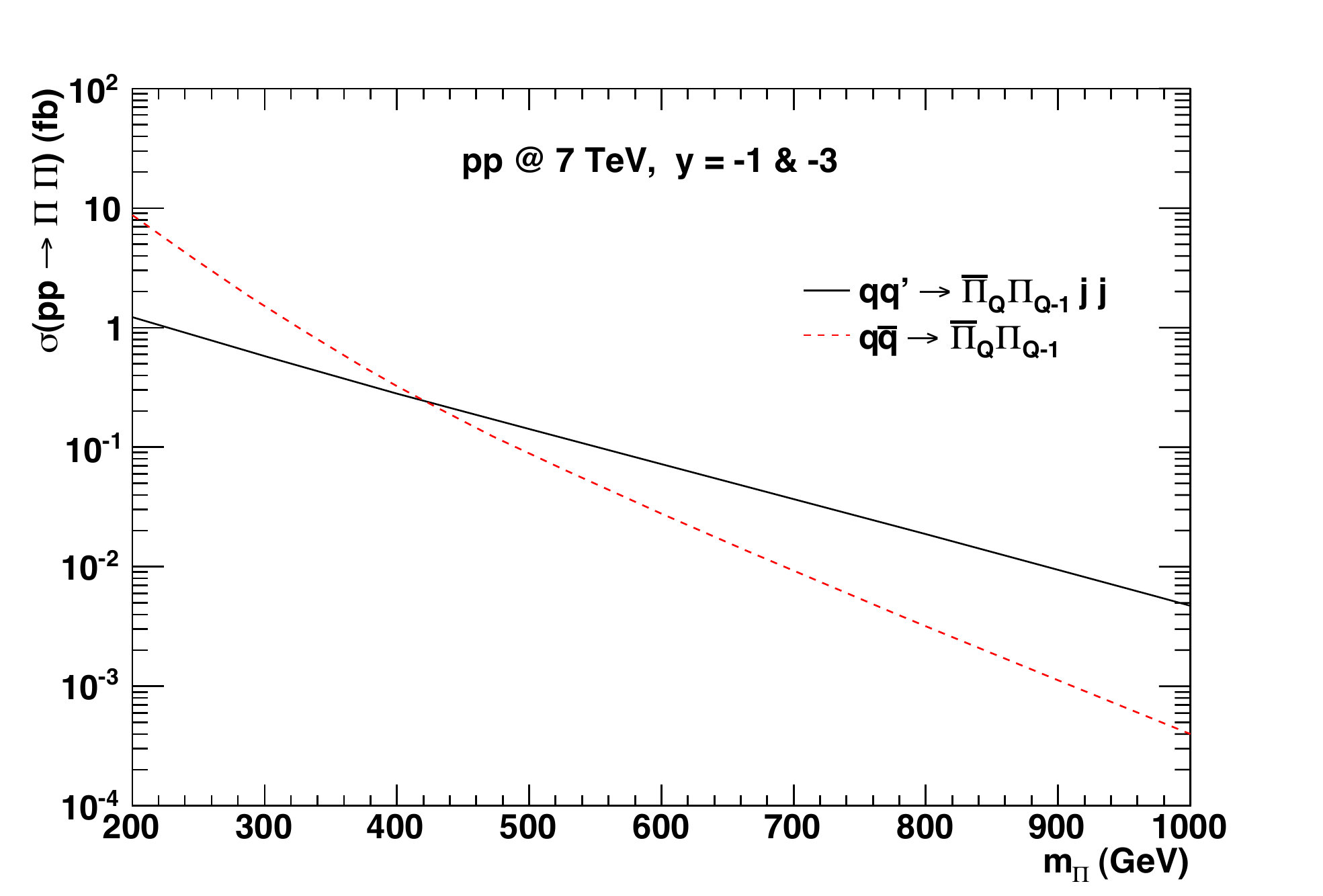}
\caption{The total cross sections for PGB production through $W$'s at the LHC as a function of  $m_{\Pi}$.} 
\label{fig:w}
\end{figure}

The cross sections for the production through the $W$ boson exchange are shown in Fig. \ref{fig:w}. Notice that we assume in this plot degenerate masses for the PGBs. With non degenerate masses the lines in Fig. \ref{fig:w} would be slightly modified according to specific final state particles.

The production cross section for a pair of PGBs of mass of around 300 GeV is $\sigma \sim 1 \textrm{ fb}$. This is much smaller than the single techinpion production of the models investigated in \cite{Farhi:1980xs,Casalbuoni:1998fs,Lane:1991qh,Lane:1999uh,Manohar:1990eg} where the typical cross section is of the oder of $\sigma \sim (10^4 -10^5) \textrm{ fb}$.

\section{PGB Decay}
\label{PGBd}
We begin by recalling that regardless of the value taken for $y$, i.e. $ \{ -3,-1,1 \}$, there will always be a doubly charged PGB 
(see table  in Section \ref{eMWT}). This fact will lead to interesting phenomenological signatures. 
Remarkably, looking at the lagrangian in \eqref{eqpieec}, the single PGB state couples only to leptons. Indeed in this section we show that a promising final state allowing to single out the MWT PGBs is the one featuring 4 charged leptons in the final state. The smoking gun final state is therefore of the type $2\ell^+ 2 \ell^-$ with equal invariant mass for each of the same sign lepton pairs. The phenomenology is similar to the one in  \cite{DelNobile:2009st}, since our PGBs, depending on the choice of $y$, resemble linear combinations of the states $\tilde{E}$, $\tilde{E}^2$ and $\tilde{E}_3$ of   \cite{DelNobile:2009st}. 
It is useful to analyze independently the two cases of $y=-3$ and $y=\pm 1$.

\subsubsection{Case $y=-3$}
With this choice of $y$, the electrical charges of $\Pi_{DD},\Pi_{UD},\Pi_{UU}$ 
are respectively $-2,-3,-4$.  The mass splitting dictated by the electroweak corrections given in \eqref{splitting} shows that the lightest PGB is the doubly charged state $\Pi_{DD}$.
For later convenience,  we will  use the symbol  
$\Pi_{Q}$ for techipions of electric charge $+|Q|$. For $y=-3$, we have $\Pi_{DD}\equiv \Pi_{2}^\dagger$. 

{Using ${\cal L}_{\Pi \psi\psi}$ given in  \eqref{eqpieec} for $y=-3$, 
one shows that this state can decay into the SM leptons via
\begin{equation}
\lambda^{ij}_2 \Pi_{2} e^c_i e^c_j + \text{h.c}, \qquad (\Pi_{2}\equiv \Pi_{DD}^\dagger)
\end{equation}
where  $i,\,j=e,\mu,\tau$. }

In this setup the remaining states $\Pi_{UD},\Pi_{UU}$ and the new leptons, once produced, are forced to
decay into the SM particles through the operator mentioned above.
\eqref{uone}  and from the requirement of absence of new stable charged particles.

We expect the direct pair production of the $\Pi_{DD}$ states, and consequent decaying topology
\begin{equation}
pp \to \Pi_{DD} \Pi_{DD}^{\dagger} \to (\ell^- \ell^-) (\ell^+ \ell^+), \qquad (\ell=e,\mu,\tau)
\end{equation}
to be the most appealing process to study experimentally. When 
the mass splitting between the PGBs is large {(i.e. $\gtrsim M_W$)}, also the final states with 5 and 6 leptons arising from the $W$ bremsstrahlung, 
\begin{equation}
\begin{split}
&pp \to  \Pi_{DD} \Pi_{UD}^{\dagger} \to \Pi_{DD} \Pi_{DD}^{\dagger} W  \to 5 \ell + \nu \\
&pp \to  \Pi_{UD} \Pi_{UD}^{\dagger} \to \Pi_{DD} \Pi_{DD}^{\dagger} W W \to 6 \ell +2\nu
\end{split}
\end{equation}
are competitive and phenomenologically interesting. For these final states, the SM background is negligible (see \cite{Akeroyd:2012nd} and references therein).  

\subsubsection{Case $y=\pm 1$}
 For $y=1$ {an interesting scenario arises due to the mixing of
 the new lepton fields and the SM ones which has been partially investigated in ~\cite{Antipin:2009ks}. Here we concentrate on complementary signatures and therefore neglect this mixing.}

{To describe the $\Pi_{UU}, \Pi_{DD}$, and $\Pi_{UD}$  decays for 
the $y=\pm 1$ cases, it is convenient 
to re-name these states in terms of their electric charge $Q=2,1,0$ 
(e.g. $\Pi_{2}, \Pi_{1}, \Pi_{0}$), 
 \begin{eqnarray}
\text{for } y=-1
\left\{
\begin{array}{l}
\Pi_{2} \equiv \Pi_{UU}^\dagger\\ 
\Pi_{1} \equiv \Pi_{UD}^\dagger\\
\Pi_{0} \equiv \Pi_{DD}^\dagger
\end{array}
\right.,\;
\text{and for } y=+1
\left\{
\begin{array}{l}
\Pi_{2} \equiv \Pi_{DD}\\ 
\Pi_{1} \equiv \Pi_{UD}\\
\Pi_{0} \equiv \Pi_{UU}
\end{array}\right.
\end{eqnarray}
Now, the PGBs interactions  with the SM fields, given in \eqref{eqpieec}, read
 \begin{eqnarray}
{\cal L}_{\Pi \psi \psi}  = 
\lambda_2^{ij}\Pi_{2}^{\dagger} e^c_i e^c_j + 
\lambda_1^{ij}\Pi_1 \nu_i e_j + 
\lambda_T^{ij}
 \begin{pmatrix}
 \nu_i, e_i
\end{pmatrix} 
 \begin{pmatrix}
 \Pi_{0} & \frac{\Pi_{1}}{\sqrt{2}} \\
\frac{\Pi_{1}}{\sqrt{2}} & \Pi_{2}
\end{pmatrix} 
\begin{pmatrix}
 \nu_j \\ 
e_j
\end{pmatrix} 
 + \textrm{ h.c.}
\label{ym1}\end{eqnarray}
Once more, the cleanest signal  
comes from $\Pi_{2}$ decaying, at the tree level, into same-sign lepton pairs. Differently from the $y=-3$ case in which only the lightest technipion could decay into SM fields, here all the states decay directly into SM fields. 
}
Here, due to the $SU(2)_L$ gauge invariance of the couplings in \eqref{ym1}, the lightest state $\Pi_{0}$
can only decay into  neutrinos. This gives rise to missing energy in
the detector, making the direct detection very difficult. The $\Pi_{1}$ decays  through the following
tree level decay processes:
\begin{equation}
\begin{split}
&\Pi_{1} \to \ell^+ \nu, \\
&\Pi_{1} \to W^+ \Pi_{0} \to  j j \nu \nu, \\
&\Pi_{1} \to W^+ \Pi_{0} \to \ell^+ \nu \nu \nu.
\end{split}
\end{equation}
If $\Pi_{1}$ is pair produced (or is produced in association with $\Pi_{2}$) then the variable $m_{T2}$ (or the transverse mass) can be used to isolate the signal. In other cases the large amount of missing energy will make the signal harder to be observed. More elaborate analysis of these different final states is beyond the scope of this study.  We will concentrate, in the following,  in analyzing the most appealing scenario corresponding to neat experimental signatures.

\section{Discovery potential and Conclusions}
\label{PGBdp}
We have constructed the low energy effective theory for the breaking pattern of $SU(4)$ to $SO(4)$ and specialized it to fit the MWT model. We have then identified the relevant decay modes of the PGBs and are now in a position to conclude this work by confronting theory and experiments. 

For every $y$, the neatest process to be investigated at the LHC involving the PGBs for which we will derive relevant constraints is
{
\begin{equation}
pp \to \Pi_{2}^{\dagger} \Pi_{2}^{} \to (\ell^-_i \ell^-_j) (\ell^+_h \ell^+_k) \qquad
\text{where } i,j,h,k=e,\mu,\tau.
\end{equation}

The flavor structure of the coupling $\lambda^{ij}\Pi_{2}l^+_i l^+_j$ depends on the ETC sector}, 
but given that this sector is unknown we will not assume any specific value for these couplings. Thus it is possible to have different lepton pairs in the final state. The effect of combining several leptonic final states to the exclusion limits is then compared against the results presented in \cite{HIG-11-007}. According to this study, the limits drawn when the doubly charged PGB is decaying into $ee$, $\mu\mu$, $e\mu$, $e\tau$ or $\mu\tau$ are comparable. The sensitivity decreases notably if the PGB decays predominantely into four $\tau$s. The authors of  \cite{DelNobile:2009st} showed that, for example, in the case of di-muons pairs in the final state, the SM background is negligible and with just few events it is possible to discover the PGBs. 

\begin{figure}[htb]
\centering
\includegraphics[width=0.7\textwidth]{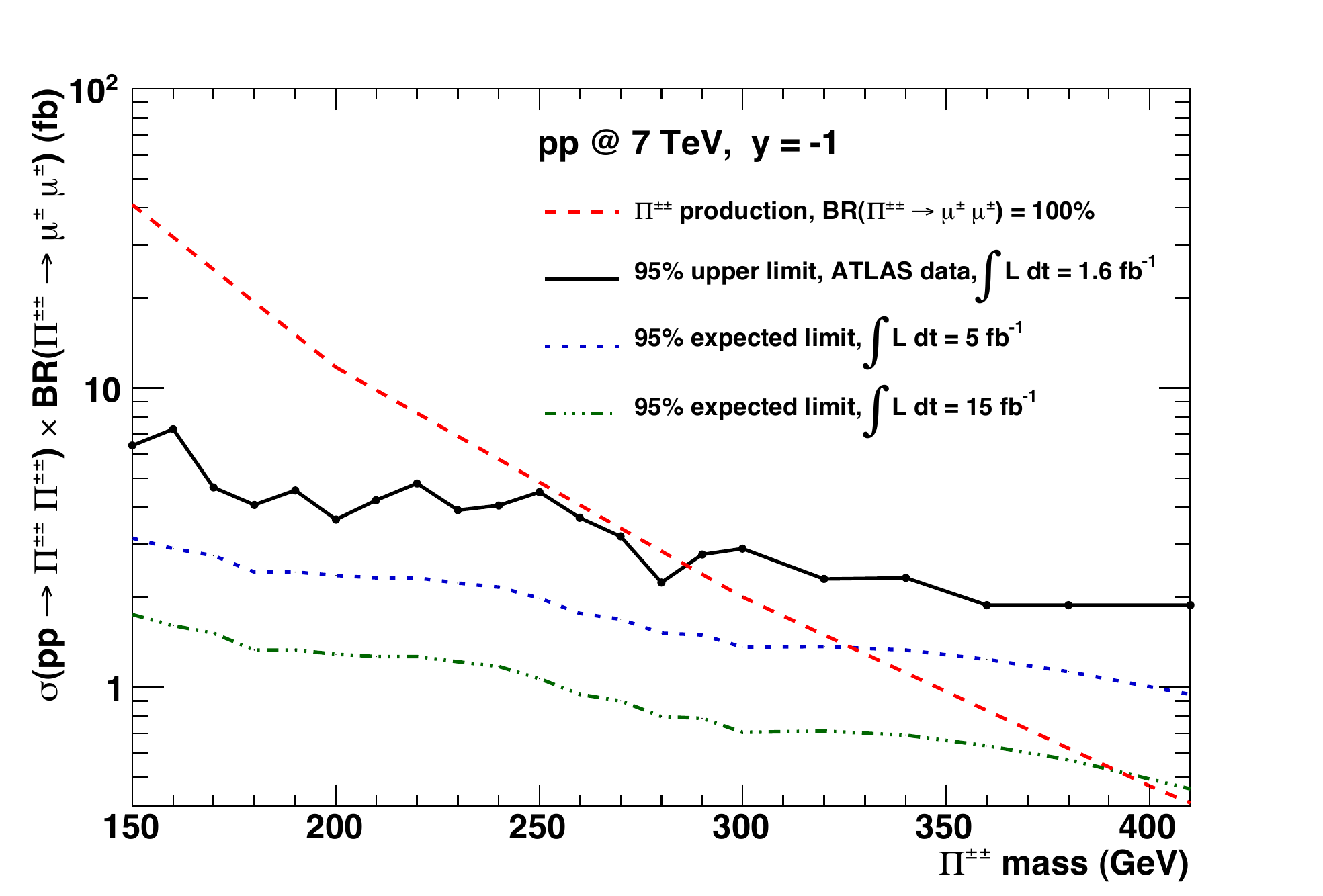}
\caption{Exclusion for  a doubly charged PGB with $y=-1$ based on the ATLAS data.} 
\label{fig:exclusion}
\end{figure}

 The ATLAS collaboration has studied the production of a doubly charged Higgs boson at the LHC in \cite{conf-127} with  1.6 fb$^{-1}$ of integrated luminosity. This study was performed by examining the invariant mass distribution of the same sign muon pairs. The doubly charged particle is assumed to decay into two muons with 100\% branching fraction. The SM background incorporates non-prompt muons from the pion and kaon decays, and prompt muons from the leptonic decays of $WW$, $WZ$ and $ZZ$. Also the production of  $t\bar{t}WW$ contributes to the background.  We have used the number of observed events and the number of expected background events reported in table 2 of \cite{conf-127} to calculated the 95\% exclusion limit for the PGB production. This yields a lower limit of 286 GeV for the PGB mass with the following acceptance cuts for the muons:
\begin{equation}
p_{\perp, \mu} > 20 GeV, \qquad |\eta_{\mu}|<2.5. 
\end{equation}
The exclusion plot is presented in Fig. \ref{fig:exclusion} for 1.6 fb$^{-1}$.  We also present the expected $95 \%$ confidence level exclusion limit for 5 and 15 fb$^{-1}$. 

From the discussion at the beginning of this section is clear that even if one allows for a more general coupling to electron and muons (but not only to taus)  the analysis above is not strongly affected.

We list below, for completeness,  several combinations of the LHC total energy and luminosity together with the doubly-charged PGB mass that can give 10 signal events.
\begin{equation}
\begin{array}{c|c|c}
\sqrt{s} & \mathcal{L}_{\textrm{int.}} & m_\Pi \\
\hline
7 \textrm { TeV} & 20 \textrm{ fb}^{-1} & 430 \textrm { GeV} \\
7 \textrm { TeV} & 100 \textrm{ fb}^{-1} & 620 \textrm { GeV} \\
8 \textrm { TeV} & 20 \textrm{ fb}^{-1} &  530 \textrm { GeV} \\
8 \textrm { TeV} & 100 \textrm{ fb}^{-1} & 750 \textrm { GeV} \\
14 \textrm { TeV} & 100 \textrm{ fb}^{-1} & 1560 \textrm { GeV} \\
14 \textrm { TeV} & 300 \textrm{ fb}^{-1} & 1880 \textrm { GeV}
\end{array}
 \end{equation}

Even though the production rates for the PGBs are smaller than the TC models featuring colored techniquarks \cite{Farhi:1980xs,Casalbuoni:1998fs,Lane:1991qh,Lane:1999uh,Manohar:1990eg}  we have shown that LHC can probe a similar mass range. This is so since in MWT the PGBs signal has negligible SM background compensating for the smaller production rates. For a direct comparison with the early models  see \cite{Chivukula:2011ue}.

We note that a doubly charged scalar appears as part of the spectrum in several extensions of the SM like the Left-Right symmetric models \cite{Azuelos:2005uc}, Little Higgs \cite{Perelstein:2005ka}, 3-3-1  \cite{CiezaMontalvo:2006zt} and  the Higgs Triplet models  \cite{Akeroyd:2012nd}.  

 \acknowledgments 
 We are happy to thank the CP$^3-$Origins colleagues 
 for fruitful discussions. We also kindly acknowledge discussions with 
 Claude Duhr, Roshan Foadi,  Mads T. Frandsen and Luca Vecchi. 
 F.M. acknowledges the
financial support from projects FPA2010-20807, 2009SGR502 and Consolider CPAN., and CSD2007-00042.
  
\newpage
\appendix

\section{SU(4) generators and PGBs quantum numbers}
\label{generators}

It is convenient to use the following representation of the $SU(4)$ hermitian generators
\beq S^a = \begin{pmatrix} \bf A & \bf B \\ {\bf B}^\dag & -{\bf A}^T
\end{pmatrix} \ , \qquad X^i = \begin{pmatrix} \bf C & \bf D \\ {\bf
    D}^\dag & {\bf C}^T \end{pmatrix} \ , \eeq
where $A$ is hermitian, $C$ is hermitian and traceless, $B = -B^T$ and
$D = D^T$. The ${S}$ are also a representation of the $SO(4)$
generators, and thus leave the vacuum invariant $SE + ES^T = 0\ $.
Explicitly, the generators read
\beq S^a = \frac{1}{2\sqrt{2}}\begin{pmatrix} \tau^a & \bf 0 \\ \bf 0 &
  -\tau^{aT} \end{pmatrix} \ , \quad a = 1,\ldots,4 \ , \eeq
where $a = 1,2,3$ are the Pauli matrices and $\tau^4 =
\mathbbm{1}$. These are the generators of $SU(2)_V \times  U(1)_V$.
\beq S^a = \frac{1}{2\sqrt{2}}\begin{pmatrix} \bf 0 & {\bf B}^a \\
{\bf B}^{a\dag} & \bf 0 \end{pmatrix} \ , \quad a = 5,6 \ , \eeq
with
\beq B^5 = \tau^2 \ , \quad B^6 = i\tau^2 \ . \eeq

Notice that $S^4, S^5$ and $S^6$ are the generators of another $SU(2)_{V'}$ algebra and that $SO(4) \simeq SU(2)_V \times SU(2)_{V'} $.

The remaining generators which do not leave the vacuum invariant are
\beq X^i = \frac{1}{2\sqrt{2}}\begin{pmatrix} \tau^i & \bf 0 \\
\bf 0 & \tau^{iT} \end{pmatrix} \ , \quad i = 1,2,3 \ , \eeq
and
\beq X^i = \frac{1}{2\sqrt{2}}\begin{pmatrix} \bf 0 & {\bf D}^i \\
{\bf D}^{i\dag} & \bf 0 \end{pmatrix} \ , \quad i = 4,\ldots,9 \ ,
\eeq
with
\beq\begin{array}{r@{\;}c@{\;}lr@{\;}c@{\;}lr@{\;}c@{\;}l}
D^4 &=& \mathbbm{1} \ , & \quad D^6 &=& \tau^3 \ , & \quad D^8 &=& \tau^1 \ , \\
D^5 &=& i\mathbbm{1} \ , & \quad D^7 &=& i\tau^3 \ , & \quad D^9 &=& i\tau^1
\ .
\end{array}\eeq

The generators are normalized as follows
\beq {\rm Tr}\left[S^aS^b\right] =\frac{1}{2}\delta^{ab}\ , \qquad \ , {\rm Tr}\left[X^iX^j\right] =
\frac{1}{2}\delta^{ij} \ , \qquad {\rm Tr}\left[X^iS^a\right] = 0 \ . \eeq

The electroweak subgroup can be embedded in $SU(4)$, as explained in detail in \cite{Appelquist:1999dq}. 

The $S^a$ generators, with $a=1,..,4$, together with the $X^a$ generators, with $a=1,2,3$, generate an $SU(2)_L \times SU(2)_R \times U(1)_V$ algebra. This is easily seen by changing genarator basis from $(S^a,X^a)$ to $(L^a,R^a)$, where
\begin{eqnarray}
L^a \equiv \frac{S^a + X^a}{\sqrt{2}} = \begin{pmatrix}\frac{\tau^a}{2}\ \ \  & 0 \\ 0 & 0\end{pmatrix} \ , \ \
{-R^a}^T \equiv \frac{S^a-X^a}{\sqrt{2}}  = \begin{pmatrix}0 & 0 \\ 0 & -\frac{{\tau^a}^T}{2}\end{pmatrix} \ ,
\end{eqnarray}
with $a=1,2,3$. The electroweak gauge group is then obtained by gauging $SU(2)_{L}$, and the $U(1)_{Y}$ subgroup of $SU(2)_R \times U(1)_V$, where
\begin{eqnarray}
Y =  -{R^3}^T + \sqrt{2}\ y \ S^4 \ .
\end{eqnarray}
The gauging of the electroweak group breaks explicitly $SU(4)$ down to  $\mathcal{H'}=SU(2)_L \times U(1)_V \times U(1)_Y$. As $SU(4)$ spontaneously breaks to $\mathcal{H}= SO(4)$, $SU(2)_L \times SU(2)_R $ breaks to $SU(2)_V$, which acts as a custodial isospin, which insures that the $\rho$ parameter is equal to one at tree-level.  

As a consequence of the explicit and spontaneous breaking of $SU(4)$, the unbroken subgroub is given by $\mathcal{H} \cap \mathcal{H'}=U(1)_{V} \times U(1)_{Q} .$ The electromagnetic group is generated by
\begin{eqnarray}
Q = \sqrt{2}\ \left( S^3 + \ y \ S^4 \right) \ .
\end{eqnarray} 
We also normalize the $U(1)_V$ generators in the following way:
\begin{eqnarray}
B_V = 2 \sqrt{2}\ S^4 \ .
\end{eqnarray}

In order to couple the PGB to the SM fermions we study the transformation properties of the matrix $U$.
Under $SU(4)$, the matrix $U$ transforms as a two index symmetric tensor, or in other words $U$ transforms  like the irreducible representation $\mathbf{10}$ of $SU(4)$.

Knowing the embedding of the electroweak generators in the $SU(4)$ algebra it is possible to decompose  the representation $\mathbf{10}$ according to the electroweak gauge group. It is useful to consider the following decomposition chain:
\begin{equation}
SU(4) \to SU(2)_L \times SU(2)_R \times U(1)_V \to SU(2)_L \times U(1)_Y \ ,
\end{equation}
according to which:
\begin{eqnarray}
\mathbf{10} &\to& (\mathbf{3},\mathbf{1})_{1} +  (\mathbf{2},\mathbf{2})_{0} +  (\mathbf{1},\mathbf{3})_{-1} \\
&\to& \mathbf{3}_{y} + \mathbf{2}_{1/2}  + \mathbf{2}_{-1/2} + \mathbf{1}_{-y+1} + \mathbf{1}_{-y}   + \mathbf{1}_{-y-1} \ .  
\end{eqnarray}
 
It is convenient to introduce the following notation:    
\begin{equation}
\begin{array}{ll}
\begin{array}{ccc}
(\mathbf{3},\mathbf{1})_{1} &\to&  \frac{T_L}{F} \equiv P_L U P_L \\
(\mathbf{2},\mathbf{2})_{0} &\to& \frac{H}{\sqrt{2} \, F} \equiv  P_L U P_R \\
(\mathbf{1},\mathbf{3})_{-1} &\to& \frac{T_R}{F}  \equiv P_R U P_R
\end{array}
&
\qquad \qquad
U= \frac{1}{F}
\begin{pmatrix}
T_L & \frac{H}{\sqrt{2}} \\
\frac{H^T}{\sqrt{2}} & T_R  
\end{pmatrix}
\end{array} \ ,
\end{equation}
as well as :
\begin{equation}
\begin{array}{cc}
\begin{array}{ccc}
\mathbf{3}_{y} &\to& \frac{T_L}{F} \equiv P_L U P_L \\
\mathbf{2}_{1/2} &\to& \frac{H_2}{\sqrt{2} \, F}\equiv P_L U P_R P_{+}  \\
\mathbf{2}_{-1/2} &\to& \frac{h_d}{\sqrt{2} \, F } \equiv P_L U P_R P_{-}  \\
\mathbf{1}_{-y-1} &\to&  \frac{S_1}{F} \equiv P_{+}  P_R U P_R P_{+}  \\
\mathbf{1}_{-y} &\to& \frac{S_2}{\sqrt{2} \, F} \equiv P_{+}  P_R U P_R P_{-}  \\
\mathbf{1}_{-y+1} &\to&  \frac{S_3}{F} \equiv P_{-} P_R U P_R P_{-}  
\end{array}
&
\qquad
U=
\frac{1}{F}
\left(
\begin{array}{cc}
T_L & 
\begin{array}{cc}
\frac{H_1}{\sqrt{2}} &
\frac{H_2}{\sqrt{2}} 
\end{array}
 \\
\begin{array}{c}
\frac{H_1^T}{\sqrt{2}} \\
\frac{H_2^T}{\sqrt{2}} 
\end{array} & 
\begin{array}{cc}
S_1 & \frac{S_2}{\sqrt{2}} \\
\frac{S_2}{\sqrt{2}} & S_3 
\end{array}
\end{array}
\right)
\end{array}
\end{equation}
with the following form for the projectors $P$:
\begin{equation}
P_L =
\begin{pmatrix}
1 & 0 & 0 & 0 \\ 
0 & 1 & 0 & 0 \\ 
0 & 0 & 0 & 0 \\ 
0 & 0 & 0 & 0 \\ 
\end{pmatrix}
\qquad
P_R =
\begin{pmatrix}
0 & 0 & 0 & 0 \\ 
0 & 0 & 0 & 0 \\ 
0 & 0 & 1 & 0 \\ 
0 & 0 & 0 & 1 \\ 
\end{pmatrix}
\qquad
P_+ =
\begin{pmatrix}
1 & 0 & 0 & 0 \\ 
0 & 0 & 0 & 0 \\ 
0 & 0 & 1 & 0 \\ 
0 & 0 & 0 & 0 \\ 
\end{pmatrix}
\qquad
P_- =
\begin{pmatrix}
0 & 0 & 0 & 0 \\ 
0 & 1 & 0 & 0 \\ 
0 & 0 & 0 & 0 \\ 
0 & 0 & 0 & 1 \\ 
\end{pmatrix}
\end{equation}

\section{Standard Model quantum numbers}
\label{SMquantum}
\begin{center}
\begin{tabular}{|c|c|c|}
\hline
Fields & $SU(2)_L$ & $U(1)_Y$ \\
\hline
$T$   & $ 3 $ & $y$ \\
$H_1$ & $2$ & $-\frac{1}{2}$ \\
$H_2$ & $2$ & $\frac{1}{2}$ \\
$S_1$ & $1$ & $1-y$ \\
$S_2$ & $1$ & $-y$ \\
$S_3$ & $1$ & $-(y+1)$ \\
\hline
$L $   & $2$ & $-\frac{3y}{2}$ \\
$E^c$ & $1$ &$\frac{3y+1}{2}$ \\
$N^c$ & $1$ & $\frac{3y-1}{2}$ \\
\hline
$q_i $   &  $2$ & $\frac{1}{6}$ \\
$u^c_i$ & $1$ & $-\frac{2}{3}$ \\
$d^c_i$ & $1$ & $\frac{1}{3}$ \\
$l_i $   & $2$ & $-\frac{1}{2}$ \\
$e^c_i$ & $1$ & $ 1 $ \\
\hline
\end{tabular}
\end{center}
with $i=1,2,3$.

\section{An effective Lagrangian for gluon fusion}
\label{eff-coup}

An effective interaction term for the PGB production via gluon fusion reads
\begin{equation}
\label{effgluon}
\mathcal{L}=-\frac{1}{4}~g(\tau)~G_{\mu\nu}^{a}G^{\mu\nu,a}\Pi_{X} \Pi^{\dagger}_{X}, \quad X =\{ UU, UD, DD\} \ .
\end{equation}
With our convention for the couplings $y_{1}$ and $y_{2}$, all the PGBs couple to the top  proportionally to its mass, similarly to the SM Higgs. Therefore the function $g(\tau)$ can be read off directly from the Higgs-gluon-gluon effective coupling upong substituting the top Yukawa coupling with \eqref{eq:top}, and replacing $m_{H}^{2}$ in the loop factor with $(p+q)^{2}$, where $p$ and $q$ are the momenta of the final state PGBs. We have $\tau=\frac{4 m_{t}^{2}}{(p+q)^{2}}$  and
\begin{equation}
g(\tau)=-\frac{i}{2 F}\frac{\alpha_{s}\sqrt{\sqrt{2}G_{F}}}{3\pi}n_{q}\left|\frac{1}{2}\tau(1+(1-\tau)f(\tau))\right|,
\end{equation}
where
\begin{equation}
f(\tau)=
\begin{cases}
-\frac{1}{4}\left( -i \pi +\log \frac{1+\sqrt{1-\tau}}{1-\sqrt{1-\tau}} \right)^{2},\quad &\rm{if} \quad \tau\ge1 \\
\left( \sin^{-1}\left( 1/\sqrt{\tau} \right) \right)^{2}, \quad &\rm{if} \quad \tau < 1
\end{cases}
\end{equation}
Of course $\alpha_s$ and $G_F$ are respectively the strong and the Fermi coupling constant.

\end{document}